\documentclass[letterpaper]{article} 
\usepackage{aaai24}  
\usepackage{times}  
\usepackage{helvet}  
\usepackage{courier}  
\usepackage[hyphens]{url}  
\usepackage{graphicx} 
\usepackage{natbib}  
\usepackage{caption} 
\usepackage{algorithm}
\usepackage{algpseudocode}

\usepackage{xcolor}
\usepackage{amsfonts}
\usepackage{amstext}
\usepackage{amsmath}
\usepackage{subfiles}
\usepackage{subfigure}
\usepackage{booktabs}
\usepackage{colortbl}
\usepackage{multirow}
\usepackage{bbding}
\usepackage{float}

\usepackage{newfloat}
\usepackage{listings}
\DeclareCaptionStyle{ruled}{labelfont=normalfont,labelsep=colon,strut=off} 
\floatstyle{ruled}
\newfloat{listing}{tb}{lst}{}
\floatname{listing}{Listing}
\title{PSC-CPI: Multi-Scale Protein Sequence-Structure Contrasting for Efficient and Generalizable Compound-Protein Interaction Prediction}

\author{
Lirong Wu{$^{1,2}$}, Yufei Huang{$^{1,2}$}, Cheng Tan{$^{1,2}$}, Zhangyang Gao{$^{1,2}$}, Bozhen Hu{$^{1,2}$}, Haitao Lin{$^{1,2}$}, \\ Zicheng Liu{$^{1,2}$}, Stan Z. Li{$^{1,^\dagger}$}\\
}
\affiliations{
$^1$ AI Lab, Research Center for Industries of the Future, Westlake University, Hangzhou, China, 310030 \\ 
$^2$ Zhejiang University, Hangzhou, China, 310058 \\
\{wulirong, huangyufei, tancheng, gaozhangyang, linhaitao, hubozhen, liuzicheng, stan.zq.li\}@westlake.edu.cn}

\usepackage{bibentry}

\begin{document}

\maketitle

\begin{abstract}
Compound-Protein Interaction (CPI) prediction aims to predict the \emph{pattern} and \emph{strength} of compound-protein interactions for rational drug discovery. Existing deep learning-based methods utilize only the single modality of protein sequences or structures and lack the co-modeling of the joint distribution of the two modalities, which may lead to significant performance drops in complex real-world scenarios due to various factors, e.g., modality missing and domain shifting. More importantly, these methods only model protein sequences and structures at a single fixed scale, neglecting more fine-grained multi-scale information, such as those embedded in key protein fragments. In this paper, we propose a novel multi-scale \emph{\underline{P}rotein \underline{S}equence-structure \underline{C}ontrasting} framework for CPI prediction (PSC-CPI), which captures the dependencies between protein sequences and structures through both intra-modality and cross-modality contrasting. We further apply length-variable protein augmentation to allow contrasting to be performed at different scales, from the amino acid level to the sequence level. Finally, in order to more fairly evaluate the model generalizability, we split the test data into four settings based on whether compounds and proteins have been observed during the training stage. Extensive experiments have shown that PSC-CPI generalizes well in all four settings, particularly in the more challenging ``\emph{Unseen-Both}" setting, where neither compounds nor proteins have been observed during training. Furthermore, even when encountering a situation of modality missing, i.e., inference with only single-modality data, PSC-CPI still exhibits comparable or even better performance than previous approaches. 
\end{abstract}

\section{Introduction}
\begin{figure}[!htbp]
    \begin{center}
        \includegraphics[width=1.0\linewidth]{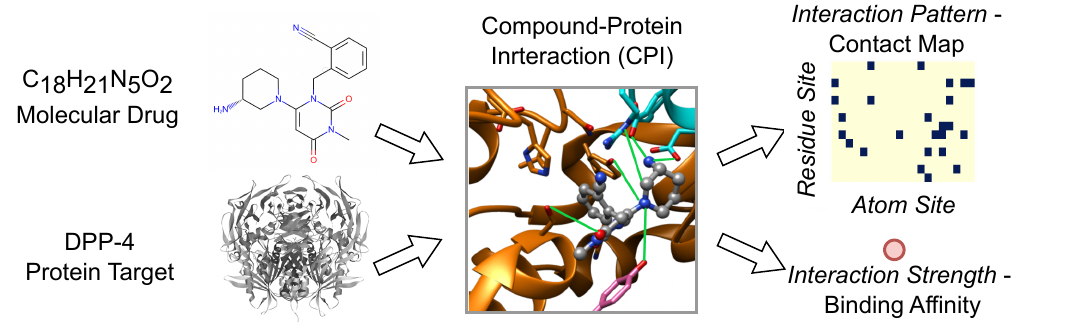}   
    \end{center}
    \caption{An illustration of Compound-Protein Interaction.}
    \label{fig:1}
\end{figure}

While various experimental assays \cite{bleicher2003hit,inglese2007high,mayr2009novel} have been applied to screen drug candidates, identifying valid drugs with desirable properties from the enormous chemical space (estimated to contain $10^{60}$ potential ``drug-like" molecule compounds  \cite{bohacek1996art,karimi2020explainable}) is still expensive and time-consuming. To overcome this bottleneck, a number of computational methods for \textit{Compound-Protein Interaction} (CPI) prediction \cite{you2020does,karimi2020explainable,gao2018interpretable,lim2021review} have been proposed to screen drugs virtually in a high-throughput way. The primary purpose of CPI prediction is to facilitate drug discovery by predicting the interaction pattern (e.g., contact map) and strength (e.g., binding affinity) of the CPI. An example of the compound-protein interaction between the protein target of dipeptidyl peptidase-4 (DPP-4) and the molecular drug of alogliptin is shown in Fig.~\ref{fig:1}.

Computational methods for CPI prediction can be mainly divided into two categories: simulation-based methods and deep learning-based methods. Molecular docking \cite{trott2010autodock,verdonk2003improved,fan2019progress,pinzi2019molecular,lin2023functional,sethi2019molecular} and molecular dynamics simulations \cite{salo2020molecular,hollingsworth2018molecular} are two typical simulation-based methods. They utilize unimodal 3D protein structures to predict both the interaction sites as well as the binding postures. Despite the remarkable success, these methods (1) rely heavily on the availability of protein 3D structures and (2) require tremendous computational resources and are very time-consuming. With the development of deep learning techniques, there have been many deep learning-based methods \cite{lim2021review} proposed for high-efficiency CPI prediction, making it possible to achieve large-scale drug screening in a relatively short time. Moreover, a large number of protein structure-free methods \cite{gao2018interpretable,li2020monn,karimi2020explainable,gao2023proteininvbench} are proposed to reduce the reliance on protein 3D structures. They can accurately predict the CPIs using only the protein sequences. However, it is the structure of a protein, rather than its sequence, that is the key to determining its functions and interactions with compounds. To combine the strengths of the two modalities, a recent work \cite{you2022cross} proposes to integrate the representations of protein sequences and structures through a complex cross-attention architecture, but it fails to model sequence-structure dependencies.

A desirable framework for CPI prediction should generally be efficient, effective, and generalizable, while two major bottlenecks deriving from real-world data may hinder the development of CPI methods.
\textbf{Modality Missing:} While joint modeling of sequence-structure is of great benefit for CPI prediction during \emph{training}, a problem often encountered in practical \emph{inference} is modality missing, i.e., there is only ONE protein modality, either sequence or structure, that is available for inference. More importantly, we cannot presuppose which modality of protein data (or both) we can obtain. \textbf{Domain Shifting:} Most existing methods work well on trainset-homologous test data but are hard to generalize to more practical (trainset-heterologous) test data, where compounds, proteins, or both have never been observed during training. Thus, how to deal with the train-test gaps in real-world scenarios is still an important issue for CPI prediction.

In this paper, we propose a novel multi-scale \emph{\underline{P}rotein \underline{S}equence-structure \underline{C}ontrasting} framework for CPI prediction (PSC-CPI) to address the above challenges. Firstly, PSC-CPI jointly pre-trains protein sequence and structure encoders to capture their dependencies by intra-modality and cross-modality contrasting. As a result, pre-trained sequence and structure encoders can enjoy the benefits of multimodal information during training, but do not require two protein modalities to be provided for inference. Secondly, a variable-length protein augmentation module is introduced, allowing both two contrasting to be performed at different scales to capture fine-grained multi-scale information embedded in key protein fragments. Finally, in order to more fairly evaluate the model generalizability, we split the test data into four settings based on whether compounds and proteins have been observed during training. Extensive experiments have shown that PSC-CPI generalizes well in all four settings, particularly in the more challenging ``\emph{Unseen-Both}" setting, where neither compounds nor proteins have been observed during training. Furthermore, PSC-CPI performs well for both unimodal and multimodal inference settings; more importantly, even when inferring with protein data of one single modality, PSC-CPI still demonstrates comparable or even better performance than previous leading methods. The source codes and related appendixes are available at: https://github.com/LirongWu/PSC-CPI.

\section{Related Work}
\subsubsection{Conventional Methods for CPI.} Identifying compound-protein interactions plays a very important role in drug discovery. Since it is expensive and time-consuming to screen drug candidates from a large chemical space through various experimental assays \cite{bleicher2003hit,inglese2007high,mayr2009novel}, virtual screening by molecular docking \cite{trott2010autodock,fan2019progress,sethi2019molecular} or molecular dynamics simulations \cite{salo2020molecular,hollingsworth2018molecular} has been studied for decades with great success in drug discovery. However, these simulation-based methods may not work well when the 3D structure of the protein is unknown or the number of known ligands is too small \cite{chen2020transformercpi}. Recent advances in deep learning have provided new insights to reduce the reliance on 3D protein structures and to develop deep learning-based methods for CPI prediction. 

\subsubsection{Deep learning-based Methods.} Most deep learning-based methods treat compounds as 1D sequences or molecular graphs and treat proteins as 1D sequences and then jointly perform representation learning and interaction prediction in an end-to-end unified framework. For example, DeepDTA \cite{ozturk2018deepdta} and DeepConvDTI \cite{lee2019deepconv} apply Convolutional Neural Networks (CNNs) \cite{lecun1995convolutional} to extract low-dimensional representations of compounds and proteins, concatenated them, and pass it into fully connected layers to predict interactions. Similarly, GraphDTA \cite{nguyen2021graphdta} treats compounds as molecular graphs and uses Graph Neural Networks (GNNs) \cite{kipf2016semi,wu2021graphmixup,wu2022knowledge,wu2023quantifying} instead of CNNs to learn compound representations. Besides, Recurrent Neural Networks (RNN) \cite{armenteros2020language} are used by DeepAffinity+ \cite{karimi2020explainable} to extract representations from sequential compounds. To better integrate compound and protein representations, TransformerCPI \cite{chen2020transformercpi} and HyperattentionDTI \cite{zhao2022hyperattentiondti} propose to learn joint compound-protein representations using a self-attentive mechanism. Recently, PerceiverCPI \cite{nguyen2023perceiver} proposes a cross-attention mechanism to improve the learning ability of the representation of compounds and protein interactions. Despite the great success, the above works have mostly modeled only the sequence information of proteins through CNN, RNN, LSTM \cite{hochreiter1997long}, etc. However, it is the structure of a protein, not the sequence, that determines its functions and interactions with compounds. For this reason, an elaborate Cross-Interaction architecture is proposed in \cite{you2022cross}, which improves CPI predictions by integrating the representations of protein sequences and structures. However, it fails to capture the sequence-structure dependencies and works only when both modalities are provided for inference. 

\subsubsection{Contrastive Learning on Proteins.}
Recent years have witnessed the great success of Contrastive Learning (CL) in protein representation learning \cite{wu2022survey,huang2023protein,huang2023data,tan2023global}. However, most previous studies have focused on contrasting within a single protein modality, either sequence \cite{lu2020self} or structure \cite{hermosilla2022contrastive,zhang2022protein}. For example, Contrastive Predictive Coding (CPC) \cite{lu2020self} applies different augmentation transformations on the input sequence to generate different views, and then maximizes the agreement of two jointly sampled pairs against that of two independently sampled pairs. In addition, Multiview Contrast \cite{hermosilla2022contrastive} proposes to randomly sample two sub-structures from each protein, encoder them into two representations, and finally maximize the similarity between representations from the same protein while minimizing that of different proteins. Despite the great progress in single-modality contrasting, relatively little work is devoted to cross-modality contrasting learning on proteins.

\begin{figure*}[!tbp]
    \begin{center}
        \includegraphics[width=0.85\textwidth]{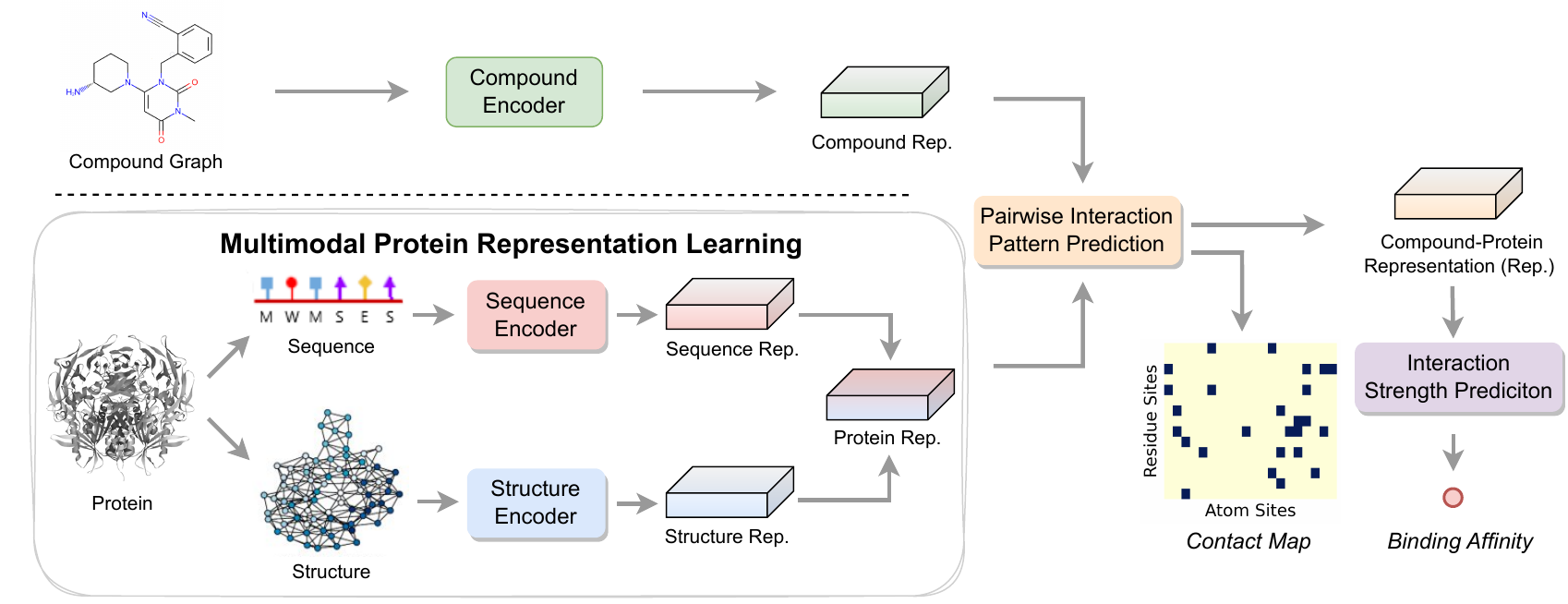}   
        \end{center}
        \caption{A high-level illustration of multi-scale protein sequence-structure contrasting framework for CPI prediction.}
\label{fig:2}
\end{figure*}

\section{Methodology}
A chemical compound can be represented by a molecular graph $\mathcal{G}_C = (\mathcal{V}_C, \mathcal{E}_C$), where each node
$a_i\in\mathcal{V}_C$ denote an atom in the compound, and each edge $e_{i,j}\in\mathcal{E}_C$ denotes a chemical bond between atom $a_i$ and atom $a_j$. A protein with $N_P$ amino acid residues can be denoted by a string of its sequence, $S=(r_1,r_2,\cdots,r_{N_P})$, where each residue $r_i$ is one of the 20 amino acid types. The amino acid sequence $S$ of a protein can be folded into a stable structure $\mathcal{G}_P$, forming a special kind of multimodal data $\mathcal{P}=(S, \mathcal{G}_P)$. The protein structure can be modeled as a protein graph $\mathcal{G}_P = (\mathcal{V}_P, \mathcal{E}_P$), where $\mathcal{V}_P$ is the node set of $N_P$ residues, and $\mathcal{E}_P \!\in\! \mathcal{V}_P \times \mathcal{V}_P$ is the set of edges that connects the residues. Given $N$ proteins $\{\mathcal{P}^{(i)}\!=\!(S^{(i)},\mathcal{G}_P^{(i)})\}_{i=1}^N$ and $M$ compounds $\{\mathcal{G}_C^{(j)}\}_{j=1}^M$, the compound-protein interaction prediction aims to learn two mappings, $\mathcal{G}_C \times \mathcal{P} \rightarrow [0,1]^{N_C \times N_P}$ and $\mathcal{G}_C \times \mathcal{P} \rightarrow \mathbb{R}_{\geq 0}$, that predict the interaction pattern and interaction strength between compounds and proteins, respectively.

\subsection{A General Framework for CPI Prediction}
A general CPI prediction framework consists of four main components: (1) \emph{compound encoder} for extracting compound representations from given compound graphs, (2) multimodal \emph{protein representation learning} for extracting protein representations from given protein sequences, structures, or both two modalities, (3) \emph{pairwise interaction pattern prediction} for predicting the contact maps between residues of a protein and atoms of a compound and learning compound-protein joint representations, and (4) \emph{interaction strength prediction} for predicting the compound-protein binding affinity. A high-level overview of the proposed PSC-CPI framework is illustrated in Fig.~\ref{fig:2}. Next, we introduce key components (1)(3)(4) and defer the discussions of multimodal protein representation learning until next section.

\subsubsection{Compound Encoder.} 
The compound encoder takes molecular graph $\mathcal{G}_C = (\mathcal{V}_C, \mathcal{E}_C)$ as input and learns a $F$-dimensional node representation for each atom. In this paper, we adopt Graph Convolutional Networks (GCNs) as the compound encoder, which is a powerful variant of GNNs that have been widely used as a feature extractor for various graph data. Given a graph $\mathcal{G}_C = (\mathcal{V}_C, \mathcal{E}_C)$, GCNs take its adjacency matrix $\mathbf{A}_C$ and node features $\mathbf{X}_C$ as input and output representation for each node. In this paper, we consider a $3$-layer GCN, which can be formulated as follows,
\begin{equation}
\mathbf{Z}^{\text{comp}}=\widehat{\mathbf{A}}\sigma\left(\widehat{\mathbf{A}} \sigma\left(\widehat{\mathbf{A}} \mathbf{X}_C \mathbf{W}^{0}\right) \mathbf{W}^{1}\right)\mathbf{W}^{2},
\label{eq:1}
\end{equation}
\noindent where $\sigma=\operatorname{ReLU}(\cdot)$, $\widehat{\mathbf{A}} = \widehat{\mathbf{D}}^{-\frac{1}{2}}(\mathbf{A}_C+\mathbf{I}) \widehat{\mathbf{D}}^{-\frac{1}{2}}$ represents a normalized adjacency matrix, $\mathbf{I}$ is an identity matrix, and $\widehat{\mathbf{D}}$ is a diagonal degree matrix for $(\mathbf{A}_C+\mathbf{I})$. In addition, $\mathbf{W}^{0} \in \mathbb{R}^{d \times F}$, $\mathbf{W}^{1} \in \mathbb{R}^{F \times F}$, and $\mathbf{W}^{2} \in \mathbb{R}^{F \times F}$ are three parameter matrices with the hidden dimension of $F$.

\subsubsection{Pairwise Interaction Pattern Prediction.} 
This module takes as inputs compound representations $\mathbf{Z}^{\text{comp}}$ and protein representations $\mathbf{Z}^{\text{prot}}$, first transforms them into a low-dimensional latent space by two independent linear transformations $\mathbf{W}^{\text{comp}}$ and $\mathbf{W}^{\text{prot}}$, then computes the interaction intensity by inner product for each residue-atom pair, and finally normalize it to obtain the interaction intensity $\mathbf{P}^{\text{cont}}[m,n]$ between $m$-th atom and $n$-th residue, as follows
\begin{equation}
\begin{small}
\begin{aligned}
\mathbf{P}^{\text{cont}}&[m,n]=\frac{\mathbf{P}^{\prime}[m,n]}{\sum_{i,j}\mathbf{P}^{\prime}[i,j]}, \text{where} \\ \mathbf{P}^{\prime} = \operatorname{Sigmoid}&\Big(\big(\sigma(\mathbf{Z}^{\text{comp}})\mathbf{W}^{\text{comp}}\big)\big(\sigma(\mathbf{Z}^{\text{prot}}) \mathbf{W}^{\text{prot}}\big)^{T}\Big).
\end{aligned}
\end{small}
\label{eq:2}
\end{equation}
To obtain the compound-protein joint embeddings, we calculate the Manhattan product of representations of each residue-atom pair and add them with $\mathbf{P}^{\text{cont}}$ as weights,
\begin{equation}
\mathbf{z}^{\text{joint}} = \sum_{m,n} \Big(\mathbf{P}^{\text{cont}}[m,n] \cdot \big(\mathbf{Z}^{\text{comp}}_m \odot \mathbf{Z}^{\text{prot}}_n\big)\Big) \in \mathbb{R}^F,
\label{eq:3}
\end{equation}
\noindent where $\mathbf{Z}^{\text{comp}}_m$  and $\mathbf{Z}^{\text{prot}}_n$ are representations of the $m$-th atom in the compound and $n$-th residue in the protein.

\begin{figure*}[!tbp]
\begin{center}
\includegraphics[width=1.0\textwidth]{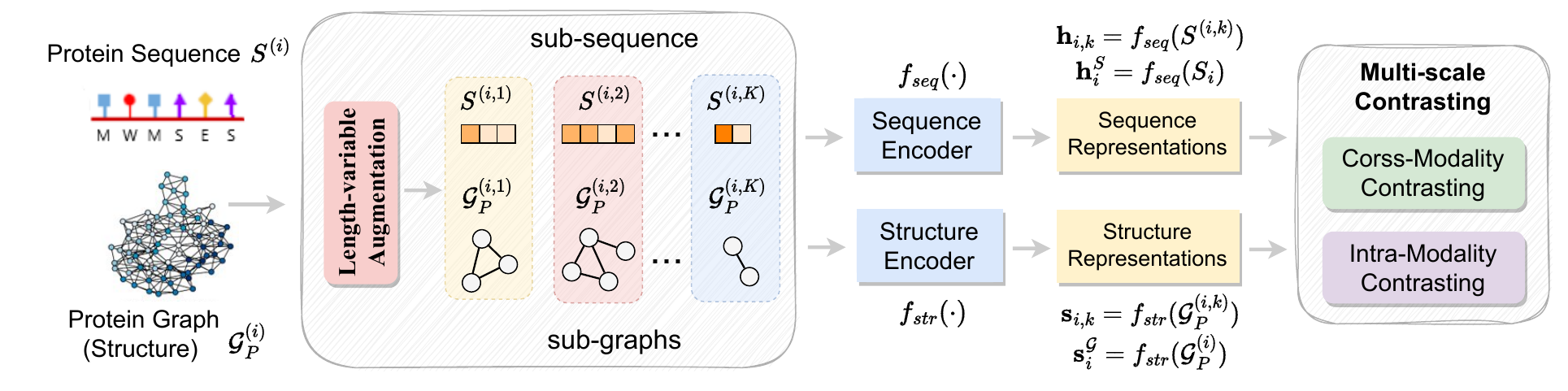}   
\end{center}
\caption{Illustration of multi-scale protein sequence-structure contrastive framework, where a length-variable augmentation module is used to generate subsequences $\{\mathcal{S}^{(i,k)}\}_{k=1}^K$ of different lengths and corresponding subgraphs $\{\mathcal{G}_P^{(i,k)}\}_{k=1}^K$, which are then encoded separately by sequence and structure encoders to perform intra- and cross-modality contrasting at different scales.}
\label{fig:3}
\end{figure*}

\subsubsection{Interaction Strength Prediction.} We take their joint embeddings $\mathbf{z}^{\text{joint}}$ as input and map it from a high-dimensional space to a non-negative value $y^{\text{aff}}$, i.e., the binding affinity. Specifically, this module consists of a layer of 1D convolution $\operatorname{Conv}(\cdot)$, a layer of maximum pooling $\operatorname{Max}(\cdot)$, and a 3-layer multilayer perceptrons $\operatorname{MLP}(\cdot)$, formulated as
\begin{equation}
y^{\text{aff}} = \operatorname{MLP}\big(\operatorname{Max}\big(\operatorname{Conv}(\mathbf{z}^{\text{joint}})\big)\big) \in \mathbb{R}_{\geq 0}.
\label{eq:4}
\end{equation}

\subsection{Multi-Scale Sequence-Structure Contrasting}
This paper aims to design an architecture-agnostic framework that is applicable to a variety of sequence and structure encoders. More importantly, we expect this framework to well handle the \emph{modality missing} problem during inference, i.e., to work well regardless of whether the protein sequence, the structure, or both modalities are provided for the inference. To achieve this, a multi-scale sequence-structure contrasting framework is proposed, as shown in Fig.~\ref{fig:3}, which fully captures the sequence-structure dependencies and multi-scale information through length-variable protein augmentation and intra-/cross-modality contrasting.

\subsubsection{Length-Variable Protein Augmentation.}
Data augmentation plays a very important role in the common contrastive learning frameworks \cite{wu2021self,he2020momentum,gao2022pifold,devlin2018bert,radford2019language}. The main purpose of data augmentation is to generate different augmented views that share the same or similar semantics as the original one. The two main challenges for data augmentation on proteins are: (1) \emph{length diversity}, different proteins may have different sequence lengths, and (2) \emph{key segment variability}, key fragments on different proteins may have very different lengths and be located at different positions on the sequence. To tackle these challenges, we augment protein data by sampling length-variable consecutive segments (subsequences) from the entire protein sequence and extracting the corresponding subgraphs. Traditional augmentation methods generally sample protein subsequences with \emph{the same length or length ratio} and then fix them before training. In this paper, we generate augmented subsequences $\{\mathcal{S}^{(i,k)}\}_{k=1}^K$ of different lengths and corresponding subgraphs $\{\mathcal{G}_P^{(i,k)}\}_{k=1}^K$ for each protein $\mathcal{P}=(\mathcal{S}^{(i)}, \mathcal{G}_P^{(i)})$ in each training epoch. As training proceeds, the length of the augmented protein subsequences keeps changing, enabling the model to \emph{``see" the same protein at more different scales}, thus capturing more multi-scale information in the protein.

\begin{figure*}[!tbp]
	\begin{center}
		\subfigure[Pre-training with protein sequence-structure pairs.]{\includegraphics[width=0.4\linewidth]{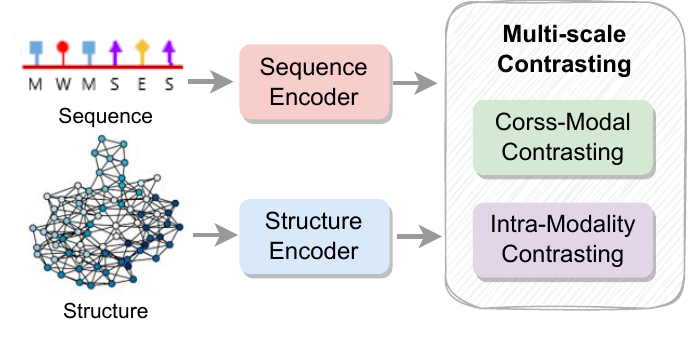} \label{fig:4a}}
		\subfigure[Inference with protein sequences and structures.]{\includegraphics[width=0.42\linewidth]{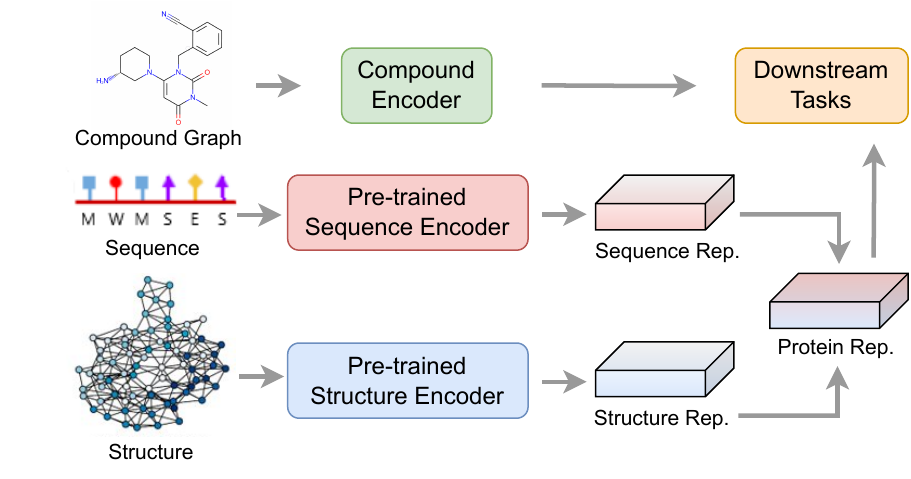} \label{fig:4b}}
		\subfigure[Inference with only protein sequences.]{\includegraphics[width=0.42\linewidth]{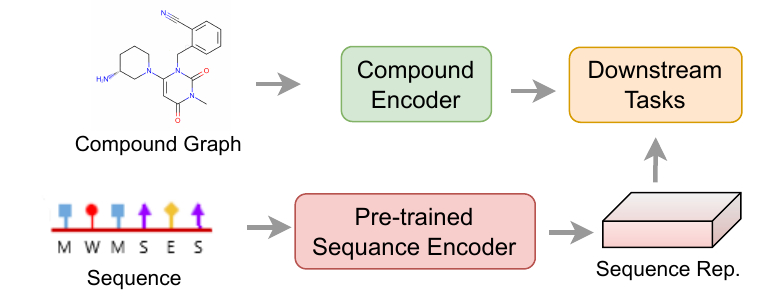} \label{fig:4c}}
		\subfigure[Inference with only protein structures.]{\includegraphics[width=0.42\linewidth]{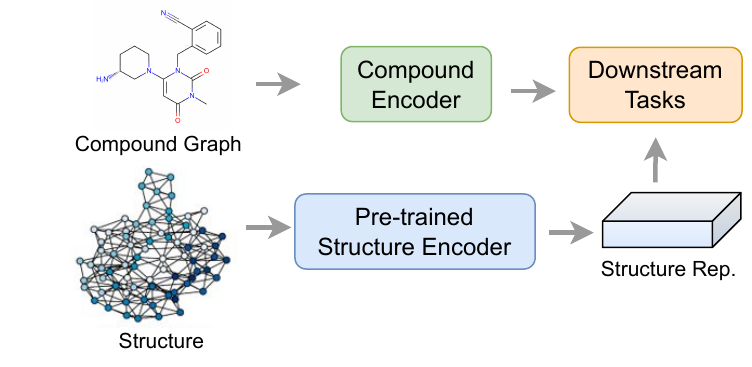} \label{fig:4d}}
	\end{center}
	\caption{\textbf{\emph{(a)}} Pre-training on known protein sequence-structure pairs by multi-scale contrasting. \textbf{\emph{(b)(c)(d)}} Three inference settings where only protein sequences, structures, or both modalities are provided.}
	\label{fig:4}
\end{figure*}

\subsubsection{Intra- and Cross-modality Contrasting.}
Two different contrastive learning objectives, i.e., intra-modality contrasting and cross-modality contrasting, are introduced in our framework to capture multi-scale information within protein sequences or structures and cross-modality dependencies. Firstly, we feed the $i$-th protein sequence $S^{(i)}$ and the augmented subsequences $\{S^{(i,k)}\}_{k=1}^K$ into a sequence encoder $f_{seq}(\cdot)$ to output the sequence representations, as follows
\begin{equation}
\mathbf{h}_i^{S} = f_{seq}(S^{(i)}), \mathbf{h}_{i,k} = f_{seq}(S^{(i,k)})
\label{eq:5}
\end{equation}
where $\  1 \leq i \leq N$, $1 \leq k \leq K$, and $K$ is the number of subsequences. Similarly, we feed the $i$-th protein graph $\mathcal{G}_P^{(i)}$ and the augmented subgraphs $\{\mathcal{G}_P^{(i,k)}\}_{k=1}^K$ into a structure encoder $f_{str}(\cdot)$ to output the structure representations,
\begin{equation}
\mathbf{s}_i^{\mathcal{G}} = f_{str}(\mathcal{G}_P^{(i)}), \mathbf{s}_{i,k} = f_{str}(\mathcal{G}_P^{(i,k)}).
\label{eq:6}
\end{equation}
Following SimCLR \cite{chen2020simple} and JOAO \cite{you2021graph}, two different two-layer MLP projection heads, denoted as $g_1(\cdot)$ and $g_2(\cdot)$, are further applied to map sequence and structure representations to a lower-dimensional space, respectively. Next, a contrastive objective function consisting of intra-modality and cross-modality contrasting is defined to pre-train the sequence and structure encoders. For intra-modality contrasting, we treat a protein sequence (protein graph) and its subsequence (subgraph) as a positive pair and subsequences from other proteins (protein graphs) in the same batch as negative pairs, which can be defined as 
\begin{equation}
\begin{small}
\begin{aligned}
\hspace{-1em}
\mathcal{L}_{\text{intra}}=\!-\!\sum_{i=1}^N\sum_{k=1}^K \bigg( \log
\underbrace{\frac{e^{\left(\operatorname{sim}\left(g_1(\mathbf{h}_i^S), g_1(\mathbf{h}_{i,k})\right) / \tau\right)}}{\sum_{b=1}^{B} e^{\left(\operatorname{sim}\left(g_1(\mathbf{h}_i^S), g_1(\mathbf{h}_{b,k})\right) / \tau\right)}}}_{\text{intra-modality (protein sequence)}} 
 \\ + \log \underbrace{\frac{e^{\left(\operatorname{sim}\left(g_2(\mathbf{s}_i^{\mathcal{G}}), g_2(\mathbf{s}_{i,k})\right) / \tau\right)}}{\sum_{b=1}^{B} e^{\left(\operatorname{sim}\left(g_2(\mathbf{s}_i^{\mathcal{G}}), g_2(\mathbf{s}_{b,k})\right) / \tau\right)}}}_{\text{intra-modality (protein graph)}}
\bigg). 
\end{aligned}
\end{small}
\label{eq:7}
\end{equation}
where $B$ is the batch size, $\operatorname{sim}(\cdot,\cdot)$ denotes the cosine similarity, and $\tau$ is the temperature coefficient. The intra-modality contrasting of Eq.~(\ref{eq:7}) transfers knowledge from protein fragments of different lengths to the final representations by maximizing the mutual information of subsequences (subgraphs) and full sequence (protein graph). Conversely, cross-modality contrasting defined in Eq.~(\ref{eq:8}) aims to capture the sequence-structure dependencies by making the subsequence and subgraph of the same protein fragment share similar semantics at different scales. Specifically, cross-modality contrasting treats subsequence and subgraph from the same protein as a positive pair, defined as
\begin{equation}
\begin{small}
\begin{aligned}
\hspace{-1em}
\mathcal{L}_{\text{cross}}=\!- \frac{1}{2}\sum_{i=1}^N\sum_{k=1}^K \bigg( \log
\frac{e^{\left(\operatorname{sim}\left(g_1(\mathbf{h}_{i,k}), g_2(\mathbf{s}_{i,k})\right) / \tau\right)}}{\sum_{b=1}^{B} e^{\left(\operatorname{sim}\left(g_1(\mathbf{h}_{i,k}), g_2(\mathbf{s}_{b,k})\right) / \tau\right)}} 
\\ + \log \frac{e^{\left(\operatorname{sim}\left(g_2(\mathbf{s}_{i,k}), g_1(\mathbf{h}_{i,k})\right) / \tau\right)}}{\sum_{b=1}^{B} e^{\left(\operatorname{sim}\left(g_2(\mathbf{s}_{i,k}), g_1(\mathbf{h}_{b,k})\right) / \tau\right)}} 
\bigg).
\end{aligned}
\end{small}
\label{eq:8}
\end{equation}
Finally, the total loss function used for the pre-trained sequence and structure encoders can be defined as
\begin{equation}
\mathcal{L}_{\text{pre}}= \mathcal{L}_{\text{cross}} + \alpha \mathcal{L}_{\text{intra}},
\label{eq:9}
\end{equation}
where $\alpha$ is a hyperparameter to trade-off between two losses.

\subsection{Training and Inference}
\subsubsection{Training.}
The CPI prediction mainly involves two downstream tasks, i.e., strength prediction and pattern prediction. When we know the ground-truth interaction strength $y^{\text{true}}_{i,j}$ between $i$-th protein and $j$-th compound, the objective function for CPI strength prediction is defined as follows,
\begin{equation}
\mathcal{L}^{\text{aff}}= \frac{1}{M\cdot N}\sum_{i=1}^N\sum_{j=1}^M\left|y^{\text{true}}_{i,j} - y^{\text{aff}}_{i,j}\right|^2.
\label{eq:10}
\end{equation}
Similarly, if we know the ground-truth interaction pattern $\mathbf{P}^{\text{true}}_{i,j}$ between $i$-th protein and $j$-th compound, the objective function for CPI pattern prediction is defined as follows,
\begin{equation}
\begin{aligned}
\mathcal{L}^{\text{cont}}= & \frac{1}{M\cdot N}\sum_{i=1}^N\sum_{j=1}^M\left\|\mathbf{P}^{\text{true}}_{i,j} - \mathbf{P}^{\text{cont}}_{i,j}\right\|_F^2 \\  + \beta & \left(\|\mathbf{P}^{\text{cont}}_{i,j}\|_{\text{group}} + \|\mathbf{P}^{\text{cont}}_{i,j}\|_{\text{fused}} + \|\mathbf{P}^{\text{cont}}_{i,j}\|_1\right),
\end{aligned}
\label{eq:11}
\end{equation}
where $\|\mathbf{P}^{\text{cont}}_{i,j}\|_{\text{group}}$ \cite{scardapane2017group}, $\|\mathbf{P}^{\text{cont}}_{i,j}\|_{\text{fused}}$ \cite{tibshirani2005sparsity}, and $\|\mathbf{P}^{\text{cont}}_{i,j}\|_{1}$ are three structure-aware sparsity regularization adopted by \cite{karimi2020explainable} to control the sparsity of the interaction contact map $\mathbf{P}^{\text{cont}}_{i,j}$. 

\subsubsection{Inference.}
While it is feasible to train the model using a small number of known sequence-structure pairs, it is overly demanding to acquire both the sequence of a protein and its structure for inference. The number of known protein structures is orders of magnitude lower than the size of the sequence dataset due to the challenges of experimental protein structure determination \cite{zhang2022protein}. The extreme data imbalance in the two modalities may lead to a \textit{modality missing} problem, i.e., existing works, while they may work well in one modality, are hard to extend to the other modality. For a more practical application scenario, we cannot presuppose which modality of protein data (or both) will be available, so developing a general framework suitable for both unimodal and multimodal inference is one of the contributions of this paper. In this paper, we have not directly integrated protein sequences and structures through architectural designs. As an alternative, we pre-trained sequence and structure encoders by performing cross-modality contrasting using \emph{a small number} of known sequence-structure pairs, aimed at aligning the representation space of sequences and structures. Consequently, despite \textit{``seeing"} only the protein sequence (structure), the representations output by the pre-trained sequence (structure) encoder also contain part of the structural (sequential) information. As a result, the pre-trained sequence and structure encoders enjoy the benefits of multimodal information during training, but do not require both two protein modalities to be provided for inference. 

\subsubsection{Illustrations and Pseudo-Code.} We provide in Fig.~\ref{fig:4} illustrations of the training and three inference settings. Taking multimodal inference with protein sequences and structures as an example, the pseudo-code for pre-training, fine-tuning, and inference is summarized in \textbf{Appendix A}.

\subsubsection{Time Complexity Analysis.} As PSC-CPI is architecture-agnostic, we do not discuss here the time complexity of compound encoder, protein sequence and structure encoder. The time complexity of remaining key modules in PSC-CPI is as follows: (1) Multi-scale Contrasting $\mathcal{O}(KN^2F)$; (2) Pattern Prediction $\mathcal{O}(MNF)$; and (3) Strength Prediction $\mathcal{O}(NF)$, where $F$ is the dimensions of hidden space and $K$ is the number of subsequences. The total time complexity $\mathcal{O}(KN^2F + MNF)$ is square and linear w.r.t the number of proteins $N$ and the number of compounds $M$, respectively.

\section{Experiments}
\subsection{Experimental Setups}
\begin{figure}[!htbp]
    \begin{center}
        \includegraphics[width=0.48\linewidth]{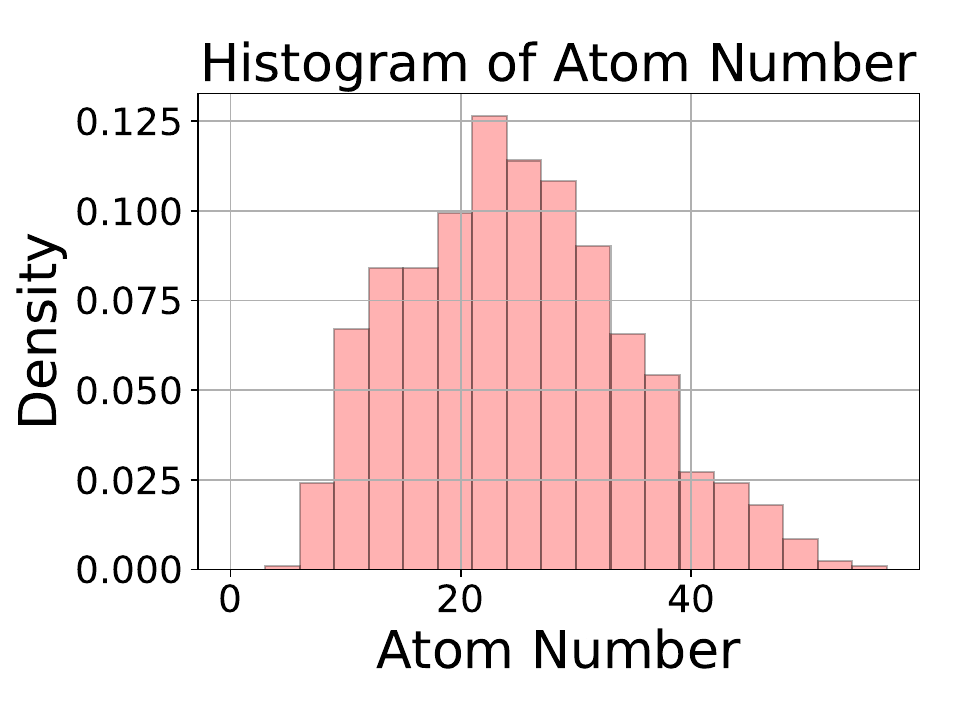}
	\includegraphics[width=0.48\linewidth]{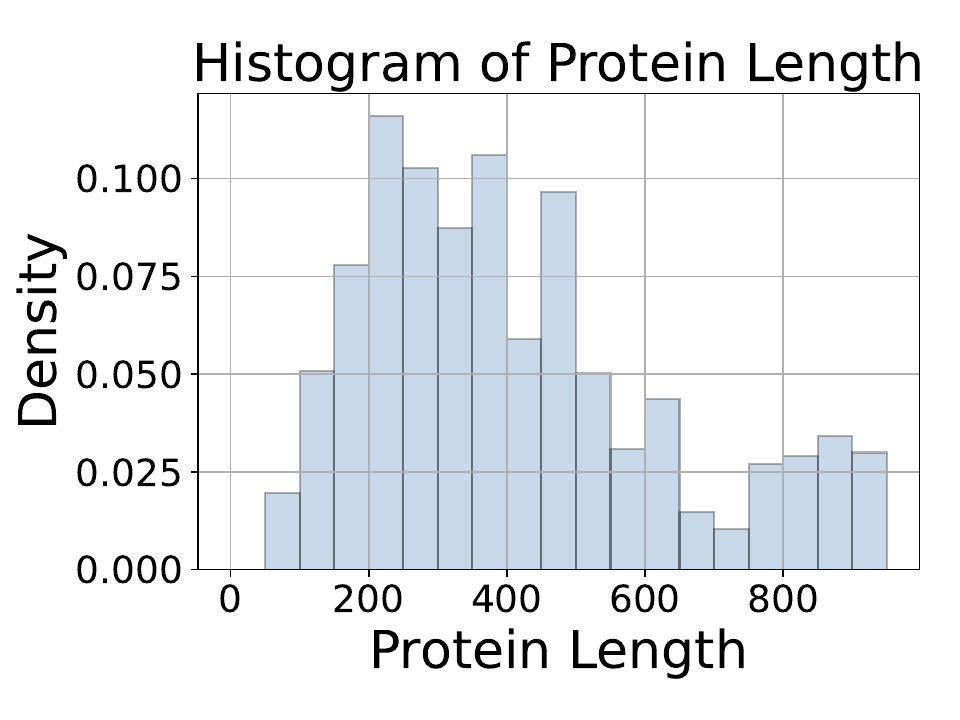}
    \end{center}
    \caption{Histogram of proteins and compounds on Karimi.}
\label{fig:5}
\end{figure}

\textbf{Datasets.} The experiments are mainly conducted on a public compound-protein dataset \cite{you2022cross,karimi2020explainable}, namely Karimi, which contains 4,446 pairs between 1,287 proteins and 3,672 compounds that are collected from PDBbind \cite{liu2015pdb} and BindingDB \cite{liu2007bindingdb}. To better evaluate the generalizability, we split the test data into four subsets based on whether compounds and proteins have been seen in the training data: (1) \emph{Seen-Both} (591 pairs): both have been seen; (2) \emph{Unseen-Comp} (521 pairs): only proteins have been seen; (3) \emph{Unseen-Prot} (795 pairs): only compounds have been seen; and (4) \emph{Unseen-Both} (202 pairs): both have never been seen. A statistical histogram of the length of the protein and the number of atoms in the compound is shown in Fig.~\ref{fig:5}. In addition, three common datasets, Davis \cite{davis2011comprehensive}, KIBA \cite{tang2014making}, and Mert \cite{metz2011navigating}, are further used to evaluate the \emph{Unseen-Both} setting, and we apply RaptorX-Contact \cite{xu2019distance} to obtain their corresponding protein graphs from protein sequences. Note that unlike previous protein pre-training methods for learning transferable knowledge from large amounts of unlabeled data, this paper aims to facilitate CPI prediction by capturing sequence-structure dependencies with a small number of known sequence-structure pairs, and thus we only pre-train on the same data provided by the downstream task without using additional unlabeled data.

\begin{table*}[!tbp]
\begin{center}

\resizebox{\textwidth}{!}{
\begin{tabular}{ccccccccccc|cc}
\toprule
\multirow{2}{*}{\textbf{Sequ.}} & \multirow{2}{*}{\textbf{Stru.}} & \multirow{2}{*}{\textbf{PSC}} & \multicolumn{2}{c}{\textbf{Seen-Both}} & \multicolumn{2}{c}{\textbf{Unseen-Comp}} & \multicolumn{2}{c}{\textbf{Unseen-Prot}} & \multicolumn{2}{c}{\textbf{Unseen-Both}} & \multicolumn{2}{c}{\textbf{Avarage}} \\ \cmidrule(r){4-6}  \cmidrule(r){6-7} \cmidrule(r){8-9} \cmidrule(r){10-11} \cmidrule(r){12-13}
 & & \multicolumn{1}{c}{} & \textbf{AUPRC} & \textbf{RMSE} & \textbf{AUPRC} & \textbf{RMSE} & \textbf{AUPRC} & \textbf{RMSE} & \textbf{AUPRC} & \textbf{RMSE} & \textbf{AUPRC} & \textbf{RMSE} \\ \midrule

\multirow{3}{*}{\CheckmarkBold} & \multirow{3}{*}{\XSolidBrush} & \XSolidBrush & 22.05 & 1.56 & 19.32 & 1.48 & 6.48 & 1.66 & 5.62 & 1.75 & 13.37 & 1.61 \\
& & \CheckmarkBold & 22.29 & 1.48 & 21.43 & 1.37 & 7.01 & 1.54 & 6.64 & 1.59 & 14.34 & 1.49 \\
& & $\Delta$ & ${\color[rgb]{0, 0, 0}\uparrow}$ 1.09\% & ${\color[rgb]{0, 0, 0}\downarrow}$ 5.13\% & ${\color[rgb]{0, 0, 0}\uparrow}$ 10.92\% & ${\color[rgb]{0, 0, 0}\downarrow}$ 7.43\% & ${\color[rgb]{0, 0, 0}\uparrow}$ 8.18\% & ${\color[rgb]{0, 0, 0}\downarrow}$ 7.23\% & ${\color[rgb]{0, 0, 0}\uparrow}$ 18.15\% & ${\color[rgb]{0, 0, 0}\downarrow}$ 9.14\% & ${\color[rgb]{0, 0, 0}\uparrow}$ 7.16\% & ${\color[rgb]{0, 0, 0}\downarrow}$ 7.45\% \\ \midrule

\multirow{3}{*}{\XSolidBrush} & \multirow{3}{*}{\CheckmarkBold} & \XSolidBrush & 22.11 & 1.58 & 21.56 & 1.52 & 10.70 & 1.73 & 9.40 & 1.80 & 15.94 & 1.66 \\
& & \CheckmarkBold & 24.26 & 1.53 & 23.78 & 1.43 & \textbf{11.14} & 1.52 & 10.62 & 1.66 & 17.45 & 1.54 \\
& & $\Delta$ & ${\color[rgb]{0, 0, 0}\uparrow}$ 9.72\% & ${\color[rgb]{0, 0, 0}\downarrow}$ 3.16\% & ${\color[rgb]{0, 0, 0}\uparrow}$ 10.30\% & ${\color[rgb]{0, 0, 0}\downarrow}$ 5.92\% & ${\color[rgb]{0, 0, 0}\uparrow}$ 4.11\% & ${\color[rgb]{0, 0, 0}\downarrow}$ 12.14\% & ${\color[rgb]{0, 0, 0}\uparrow}$ 12.98\% & ${\color[rgb]{0, 0, 0}\downarrow}$ 7.78\% & ${\color[rgb]{0, 0, 0}\uparrow}$ 9.47\% & ${\color[rgb]{0, 0, 0}\downarrow}$ 7.23\% \\ \midrule

\multirow{3}{*}{\CheckmarkBold} & \multirow{3}{*}{\CheckmarkBold} & \XSolidBrush & 23.86 & 1.55 & 23.12 & 1.44 & 9.06 & 1.61 & 8.52 & 1.65 & 16.14 & 1.56 \\
& & \CheckmarkBold & \textbf{25.42} & \textbf{1.42} & \textbf{24.67} & \textbf{1.31} & 11.03 & \textbf{1.47} & \textbf{11.65} & \textbf{1.52} & \textbf{18.19} & \textbf{1.43} \\
& & $\Delta$ & ${\color[rgb]{0, 0, 0}\uparrow}$ 6.54\% & ${\color[rgb]{0, 0, 0}\downarrow}$ 8.39\% & ${\color[rgb]{0, 0, 0}\uparrow}$ 6.70\% & ${\color[rgb]{0, 0, 0}\downarrow}$ 9.03\% & ${\color[rgb]{0, 0, 0}\uparrow}$ 21.74\% & ${\color[rgb]{0, 0, 0}\downarrow}$ 8.70\% & ${\color[rgb]{0, 0, 0}\uparrow}$ 36.73\%  & ${\color[rgb]{0, 0, 0}\downarrow}$ 7.88\% & ${\color[rgb]{0, 0, 0}\uparrow}$ 12.70\% & ${\color[rgb]{0, 0, 0}\downarrow}$ 8.33\% \\ \bottomrule
 
\end{tabular}}
\caption{Performance comparison of CPI models pre-trained w/ and w/o PSC on pattern prediction (measured by AUPRC, higher is better) and strength prediction (measured by RMSE, lower is better) under four data splits on the Karimi dataset, where the best metrics are marked in bold. ${\color[rgb]{0, 0, 0}\uparrow}$" and ${\color[rgb]{0, 0, 0}\downarrow}$ denote the gains and drops w.r.t the vanilla model w/o PSC, respectively.}
\label{tab:1}
\end{center}
\end{table*}

\begin{table*}[!htbp]
\begin{center}

\resizebox{1.0\textwidth}{!}{
\begin{tabular}{lcccccc}
\toprule
\multirow{2}{*}{\textbf{Methods}} & \multicolumn{2}{c}{\textbf{Davis}} & \multicolumn{2}{c}{\textbf{KIBA}} & \multicolumn{2}{c}{\textbf{Metz}} \\ \cmidrule(r){2-3}  \cmidrule(r){4-5} \cmidrule(r){6-7}
 & \textbf{MSE} $\downarrow$ & \textbf{CIndex} $\uparrow$ & \textbf{MSE} $\downarrow$ & \textbf{CIndex} $\uparrow$ & \textbf{MSE} $\downarrow$ & \textbf{CIndex} $\uparrow$ \\ \midrule
DeepConvDTI \cite{lee2019deepconv} & $0.598_{\pm0.057}$ & $0.546_{\pm0.043}$ & $0.550_{\pm0.009}$ & $0.635_{\pm0.007}$ & $0.703_{\pm0.027}$ & $0.671_{\pm0.016}$ \\
GraphDTA \cite{nguyen2021graphdta} & $0.846_{\pm0.058}$ & $0.459_{\pm0.032}$ & $0.698_{\pm0.042}$ & $0.591_{\pm0.013}$ & $1.232_{\pm0.094}$ & $0.615_{\pm0.010}$ \\
DeepAffinity+ \cite{karimi2020explainable} & $0.710_{\pm0.044}$ & $0.473_{\pm0.038}$ & $0.658_{\pm0.051}$ & $0.574_{\pm0.024}$ & $0.927_{\pm0.062}$ & $0.626_{\pm0.020}$ \\
HyperattentionDTI \cite{zhao2022hyperattentiondti} & $0.671_{\pm0.045}$ & $0.517_{\pm0.013}$ & $1.022_{\pm0.062}$ & $0.590_{\pm0.015}$ & $1.064_{\pm0.080}$ & $0.630_{\pm0.013}$ \\
TransformerCPI \cite{chen2020transformercpi} & $0.549_{\pm0.038}$ & $0.490_{\pm0.032}$ & $0.630_{\pm0.057}$ & $0.563_{\pm0.014}$ & $1.081_{\pm0.125}$ & $0.557_{\pm0.016}$ \\
PerceiverCPI \cite{nguyen2023perceiver} & $\underline{0.463}_{\pm0.013}$ & $\textbf{0.638}_{\pm0.028}$ & $\underline{0.522}_{\pm0.010}$ & $\underline{0.638}_{\pm0.013}$ & $0.658_{\pm0.016}$ & $\underline{0.675}_{\pm0.012}$ \\
Cross-Interaction \cite{you2022cross} & $0.514_{\pm0.037}$ & $0.586_{\pm0.040}$ & $0.558_{\pm0.028}$ & $0.618_{\pm0.021}$ & $\underline{0.642}_{\pm0.036}$ & $0.672_{\pm0.028}$ \\
PSC-CPI (ours) & $\textbf{0.455}_{\pm0.026}$ & $\underline{0.624}_{\pm0.033}$ & $\textbf{0.490}_{\pm0.018}$ & $\textbf{0.664}_{\pm0.017}$ & $\textbf{0.595}_{\pm0.024}$ & $\textbf{0.701}_{\pm0.023}$ \\ \midrule

\end{tabular}} 
\caption{Performance comparison of PSC-CPI with other state-of-the-art baselines for CPI strength prediction on three public datasets under the Unseen-Both setting, where the best and second metrics are marked as bold and underline, respectively.}
\label{tab:2}
\end{center}
\end{table*}

\noindent \textbf{Hyperparameter.} The hyperparameters are set the same for all four datasets: Adam optimizer with learning rate $lr$ = 5e-5, weight decay $decay$ = 5e-4, $\beta$ = 0.001, and Epoch $E$ = 200. The other dataset-specific hyperparameters are determined by an AutoML toolkit NNI with the search spaces as hidden dimension $F=\{64, 128, 256, 512\}$; batch size $B=\{16, 32, 64\}$, temperature $\tau=\{0.1, 0.3, 0.5, 0.8\}$, loss weight $\lambda=\{0.1, 0.3, 0.5, 0.8, 1.0\}$, and $K=\{1,3,5,10\}$.

\subsection{Comparative Results}
To evaluate the effectiveness of PSC-CPI under \textit{modality missing} during inference, we report the performance of CPI pattern prediction (measured by AUPRC) and strength prediction (measured by RMSE) under four different test data splits on the Karimi dataset in Table.~\ref{tab:1}. We first train the model using a small number of known sequence-structure pairs and then test its generalization under three different inference settings, where protein sequence, structure, or both modalities are provided. For unimodal inference with only protein sequences or structures, we adopt HRNN and GAT \cite{velivckovic2017graph} as sequence encoder and structure encoder, respectively. For multimodal inference with both sequences and structures, we concatenate the outputs of HRNN and GAT as final representations. We can observe from Table.~\ref{tab:1} that: (1) \emph{Strengths of multimodality.} The protein sequences and structures have their own strengths for different tasks; for example, sequences are more beneficial for strength prediction, while structures help more in predicting interaction patterns. However, inference with both modalities, either pre-trained w/ and w/o PSC, combines their strengths and outperforms both individual modalities. (2) \emph{Applicability to different inference settings.} The performance of pre-training with PSC consistently improves the vanilla CPI model w/o PSC regardless of the unimodal or multimodal inference. More importantly, by pre-training with PSC, the results of inference with only unimodal data can even outperform multimodal methods. (3) \emph{Generalizability.} Pre-training with PSC shows noticeable advantages under all four test data splits, especially in the \emph{``Unseen-Both"} setting. Due to space limitations, experiments on more metrics and architectures are placed in \textbf{Appendix B}.

To further compare PSC-CPI with other state-of-the-art (SOTA) competitors, we evaluated their performance of CPI binding affinity prediction on three public datasets (Davis, KIBA, and Mert, all in the ``\emph{Unseen-Both}" setting), using Mean Squared Error (MSE) and Concordance Index (CIndex) as metrics. The benchmarks for comparison include DeepConvDTI, TransformerCPI, HyperattentionDTI, PerceiverCPI, DeepAffinity+, and Cross-Interaction. Following the experimental setup in \cite{nguyen2023perceiver}, we transform a few binary classification models, such as TransformerCPI, DeepconvDTI, and HyperattentionDTI into regression models by modifying their final layers for a fair comparison. From the reported results in Table.~\ref{tab:2}, it can be observed that two multimodal methods, Cross-Interaction and PSC, both show fairly good performance. However, Cross-Interaction still slightly lags behind the SOTA method - PerceiverCPI, while our PSC-CPI exceeds PerceiverCPI by a little bit, achieving the best in 5 out of 6 metrics for the three datasets.

\begin{figure*}[!tbp]
	\begin{center}
		\subfigure[AUPRC $\sim$ Length]{\includegraphics[width=0.235\linewidth]{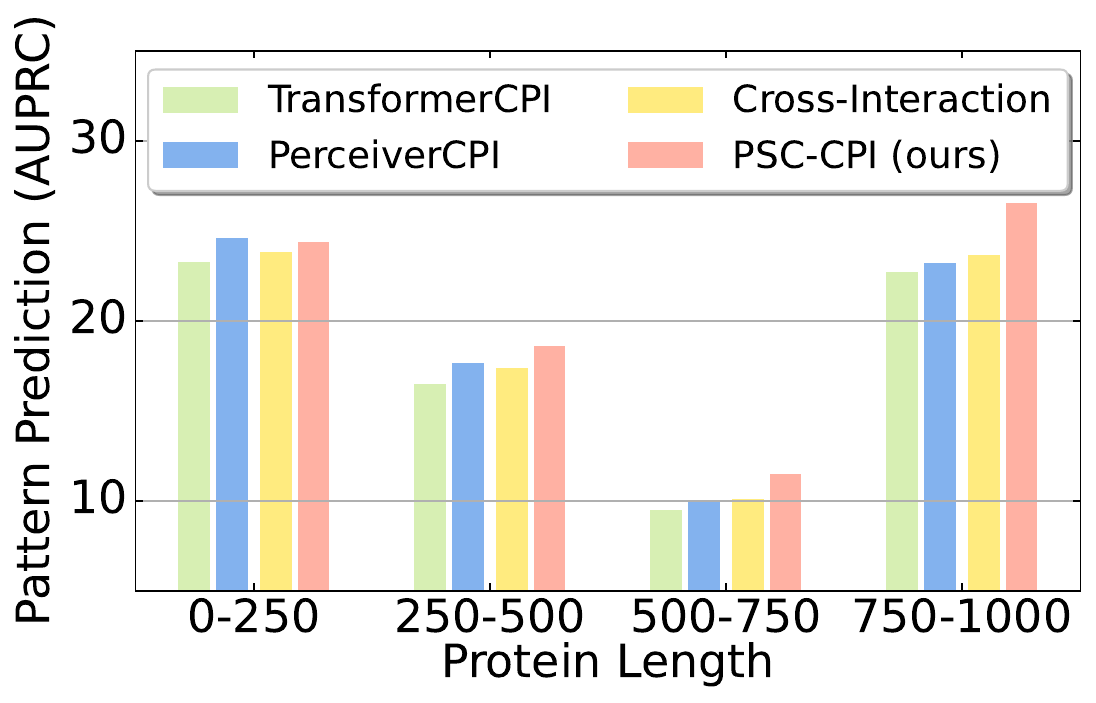} \label{fig:6a}}
		\subfigure[AUPRC $\sim$ \#Atom]{\includegraphics[width=0.235\linewidth]{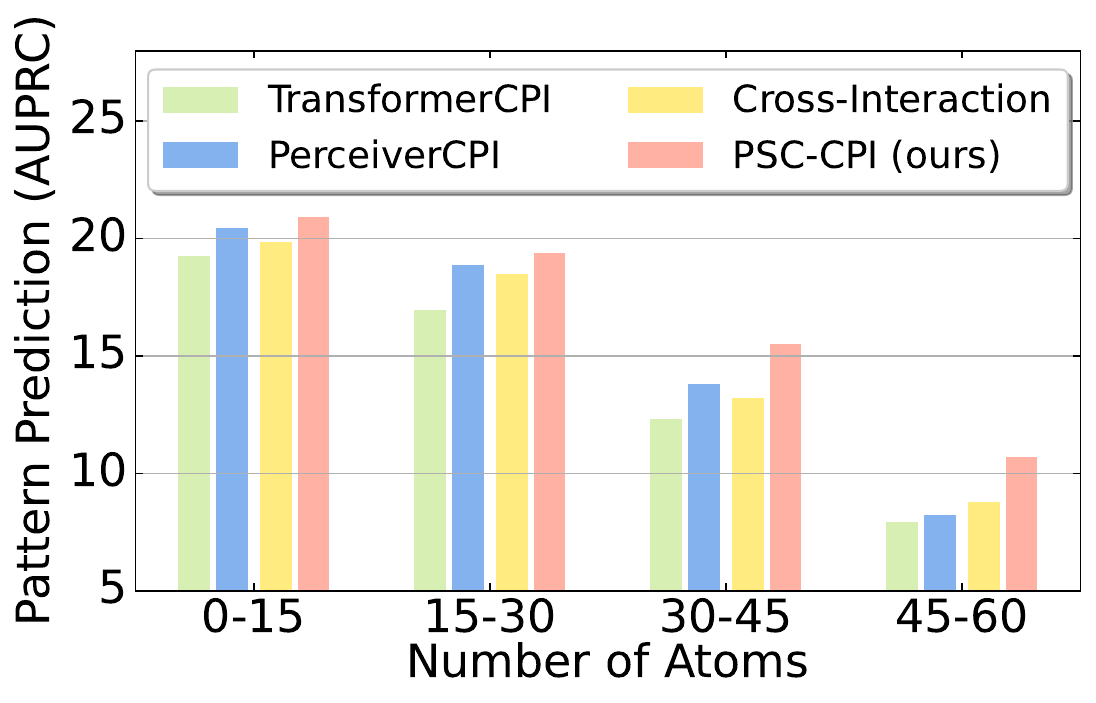} \label{fig:6b}}
		\subfigure[RMSE $\sim$ Length]{\includegraphics[width=0.235\linewidth]{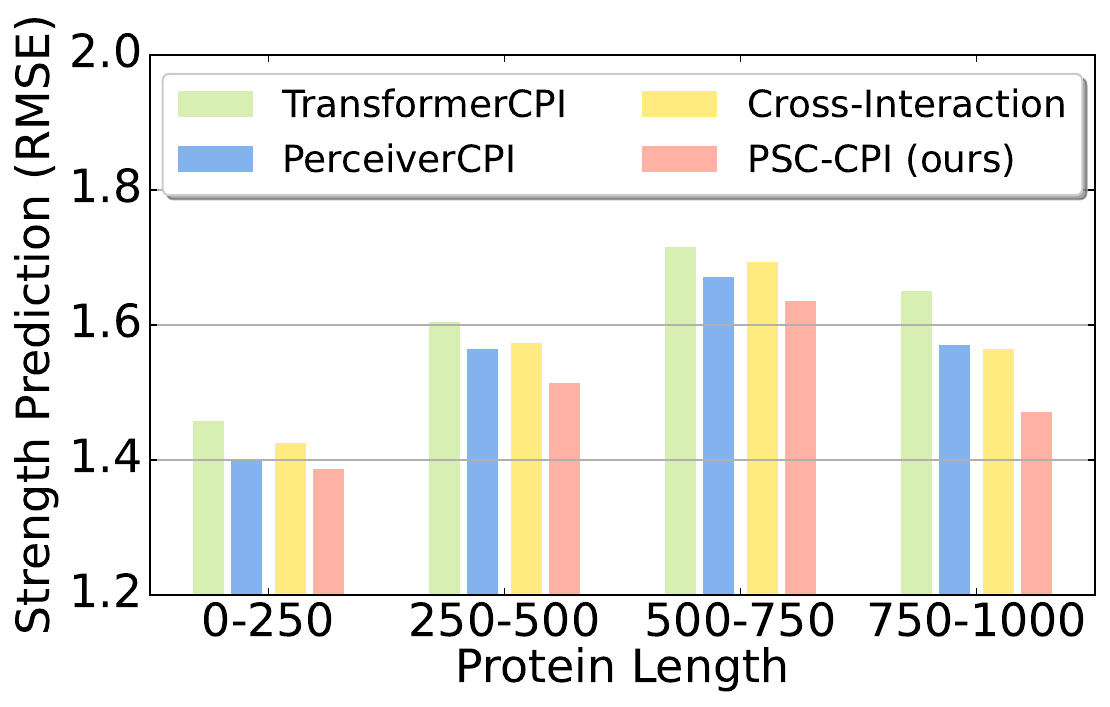} \label{fig:6c}}
		\subfigure[RMSE $\sim$ \#Atom]{\includegraphics[width=0.235\linewidth]{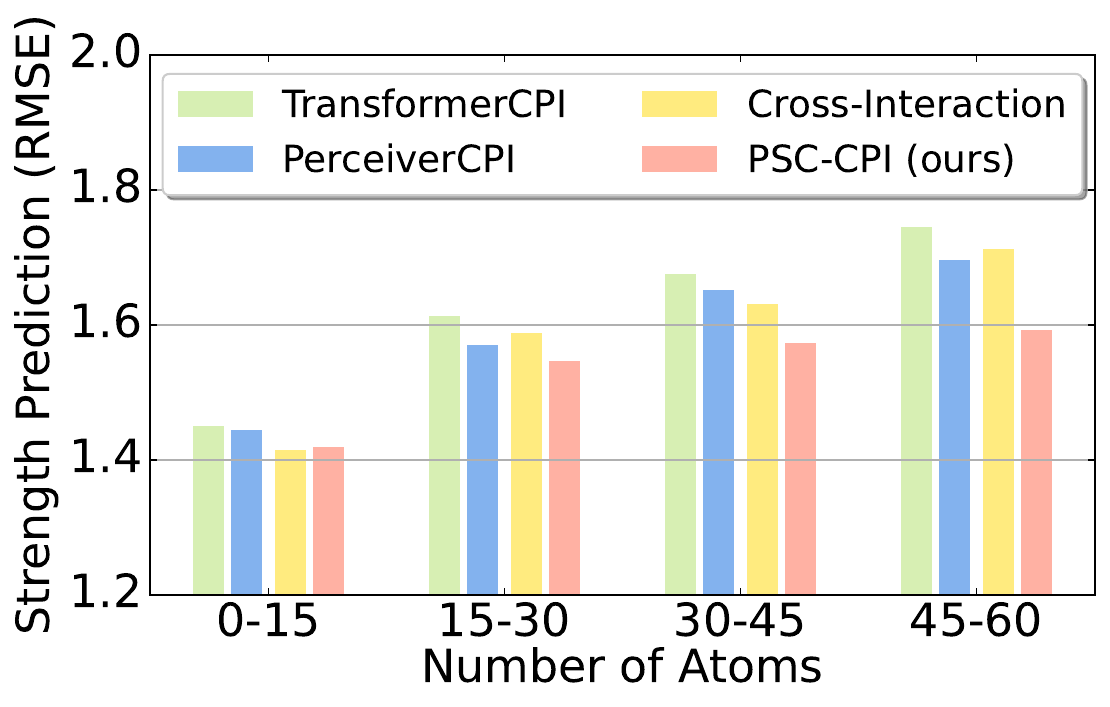} \label{fig:6d}}
	\end{center}
	\caption{Performance of four representative methods for CPI pattern prediction (measured by AUPRC, higher is better) and CPI strength prediction (measured by RMSE, lower is better) under different protein lengths and number of atoms.}
	\label{fig:6}
\end{figure*}

\begin{figure*}[!htbp]
    \begin{center}
        \includegraphics[width=1.0\textwidth]{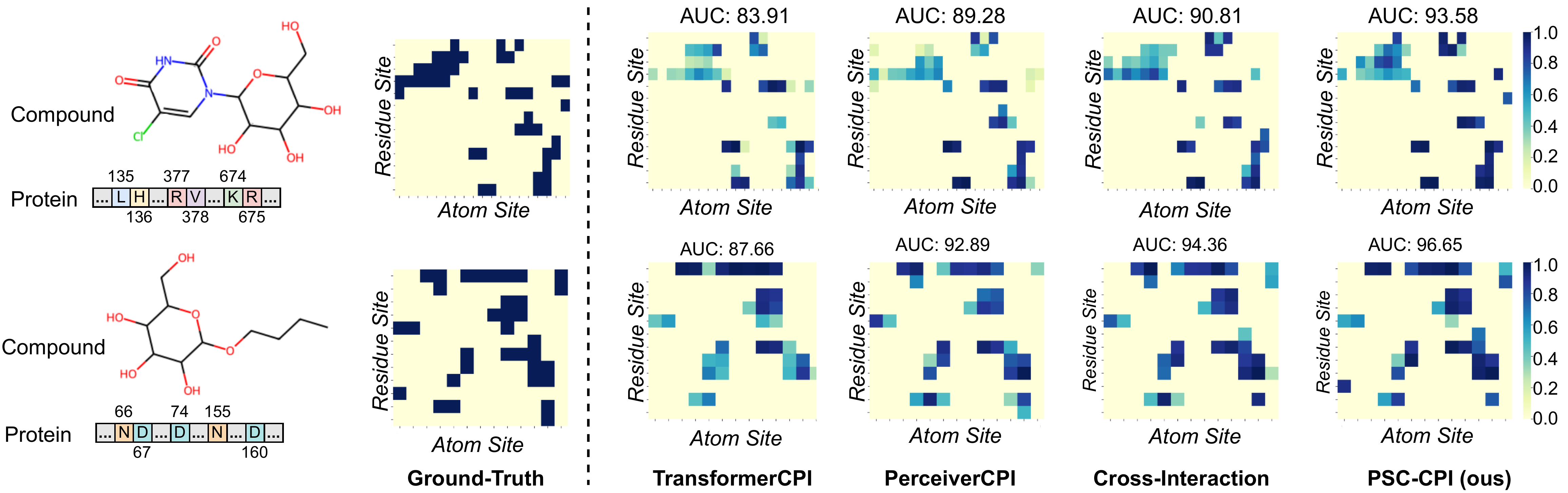}   
        \end{center}
    \caption{Visualization of predicted contact maps for various methods, along with their AUC scores.}
\label{fig:7}
\end{figure*}

\subsection{Evaluation on Protein Lengths and Atom Numbers}
To compare the performance of PSC-CPI with other baselines at different protein lengths and number of atoms, we select three representative methods (TransformerCPI, PerceiverCPI, and Cross-Interaction) and report their performance \textit{averaged over four different test data splits} on the Karimi dataset, where strength prediction and pattern prediction are measured by AUPRC (higher is better) and RMSE (lower is better), respectively. As can be seen from the results in Fig.~\ref{fig:6}, the \textbf{performance gains} of PSC-CPI over other baselines keep expanding as the protein length and number of atoms increase. This indicates that PSC-CPI has a greater advantage in dealing with complex proteins or compounds due to the multi-scale information it captures.

\begin{table*}[!tbp]
\begin{center}
\resizebox{0.8\textwidth}{!}{
\begin{tabular}{lcccccccc}
\toprule
\multicolumn{1}{c}{\multirow{2}{*}{\textbf{Methods}}} & \multicolumn{2}{c}{\textbf{Seen-Both}} & \multicolumn{2}{c}{\textbf{Unseen-Comp}} & \multicolumn{2}{c}{\textbf{Unseen-Prot}} & \multicolumn{2}{c}{\textbf{Unseen-Both}} \\ \cmidrule(r){2-3}  \cmidrule(r){4-5} \cmidrule(r){6-7} \cmidrule(r){8-9}
 & \textbf{AUPRC} & \textbf{RMSE} & \textbf{AUPRC} & \textbf{RMSE} & \textbf{AUPRC} & \textbf{RMSE} & \textbf{AUPRC} & \textbf{RMSE} \\ \midrule
Vanilla CPI (w/o PSC) & 23.86 & 1.55 & 23.12 & 1.44 & 9.06 & 1.61 & 8.52 & 1.65 \\
PSC-CPI (full model) & \textbf{25.42} & \textbf{1.42} & \textbf{24.67} & \textbf{1.31} & \textbf{11.03} & \textbf{1.47} & \textbf{11.65} & \textbf{1.52} \\
\quad w/o Intra-modality CL & \underline{25.36} & \underline{1.44} & \underline{24.47} & \underline{1.32} & \underline{10.46} & 1.50 & \underline{11.30} & \underline{1.54} \\
\quad  w/o Cross-modality CL & 25.06 & 1.47 & 23.95 & 1.36 & 10.88 & \underline{1.48} & 10.34 & 1.58 \\
\quad w/o Length-variable DA & 24.69 & 1.49 & 24.00 & 1.37 & 10.69 & 1.53 & 10.41 & 1.60 \\  \bottomrule

\end{tabular}}
\caption{Ablation study on intra- and cross-modality contrastive losses and data augmentation used for pre-training.}
\label{tab:3}
\end{center}
\end{table*}

\subsection{Ablation Study \& Visualizations}
To explore how different pre-training contrastive losses and augmentation strategies influence performance, we compare the vanilla CPI model (without pre-training with PSC) with four other schemes: (A) \textit{full model}: pre-training with both two contrasting and length-variable augmentation; (B) \textit{w/o Intra-modality CL}: pre-training without intra-modality contrasting; (C) \textit{w/o Cross-modality CL}: pre-training without cross-modality contrasting; and (D) \textit{w/o length-variable augmentation}: pre-training with both two contrasting but with \textbf{length-fixed} augmentation. We can observe from Table.~\ref{tab:3} that (1) Both intra-modality and cross-modality contrasting help improve performance, especially the latter, suggesting that the sequence-structure dependence helps more than the multi-scale information. (2) The full model combines the strengths of two contrasting methods and outperforms both. (3) Data augmentation plays a very important role in contrastive learning. While length-fixed augmentation works as well, it performs poorer than length-variable augmentation as it ignores the multi-scale information.

To visualize the interaction patterns of our PSC-CPI and other baselines, we select two representative compound-protein pairs and plot the ground-truth labels and predicted results of the contact maps between potential residue sites and atomic sites for four representative methods. In addition, we threshold the predicted contact maps to make their interaction numbers equal to the ground-truth interaction numbers between the protein and the compound, and finally normalize them by the maximum value. For a fair comparison, all four methods default to adopt the pairwise interaction pattern prediction module proposed in this paper. As can be seen from the visualizations and AUC scores in Fig.~\ref{fig:7}, PerceiverCPI and Cross-Interaction perform much better than TransformerCPI, but still lag far behind PSC-CPI in terms of both qualitative visualizations and quantitative scores.

\section{Conclusion}
In this paper, we propose a novel multi-scale \emph{\underline{P}rotein \underline{S}equence-structure \underline{C}ontrasting} (PSC) framework for CPI prediction that is applicable to inference on both unimodal and multimodal protein data. Owing to the length-variable augmentation and intra- and cross-modality contrasting, PSC-CPI has the capability to capture sequence-structure dependencies and multi-scale information, performing well for proteins of various sequence lengths and compounds of various atomic numbers. Extensive experiments show that PSC-CPI generalizes well across various data, especially in a more challenging ``\emph{Unseen-Both}" setting, where neither compounds nor proteins have observed seen during training. Despite much progress, limitations remain. For example, multi-scale modeling only involves the residue-protein scale, and it may be a promising direction to extend it to the atomic scale of proteins. Moreover, we have not explored in depth the efficiency issue, which will be left for future work.

\clearpage
\section{Acknowledgments}
This work was supported by National Key R\&D Program of China (No. 2022ZD0115100), National Natural Science Foundation of China Project (No. U21A20427), and Project (No. WU2022A009) from the Center of Synthetic Biology and Integrated Bioengineering of Westlake University.

\bibliography{aaai24}


\end{document}